# Multi-Arrival Infrasound from Meteoroids: Fragmentation Signatures versus Propagation Effects in a Fine-Scale Layered Atmosphere

Igor P. Chunchuzov[1,*], Oleg E. Popov[1], Elizabeth A. Silber[2,3] and Sergey N. Kulichkov[1,4]

[1]Obukhov Institute of Atmospheric Physics, Russian Academy of Sciences, Moscow, 119017 Russia
[2]Sandia National Laboratories, 1515 Eubank Blvd. NE, Albuquerque, NM, 87123, US
[3]Department of Earth Sciences, Western University, London, ON, N6A 5B7, Canada
[4]Moscow State University, Moscow, 119991 Russia

*Corresponding author: igor.chunchuzov@gmail.com

**Abstract**

Infrasonic signatures of meteoroid fragmentation are frequently ambiguous: do multiple arrivals signify a complex breakup or merely the distorting effects of a layered atmosphere? Resolving this ambiguity is critical for accurate energy estimates and source reconstruction. In this study, we address this challenge by analyzing a unique regional dataset of well-constrained meteoroid events observed by the Southern Ontario Meteor Network and the co-located Elginfield Infrasound Array. We employ pseudo-differential parabolic equation (PPE) simulations to quantify how fine-scale gravity-wave structures in the stratosphere and lower thermosphere modify acoustic waveforms at ranges <300 km. Our modeling reveals that while fine-scale layering can stretch signals and generate diffuse oscillatory tails, it does not produce discrete, high-amplitude pulse splitting at ranges below ~140 km. By applying these results to the rare multi-arrival event 20060305, we demonstrate that its distinct double arrival at 100 km range is inconsistent with atmospheric multipathing and provides definitive evidence of separate fragmentation episodes. These findings establish new diagnostic criteria for separating source physics from propagation artifacts, improving the reliability of infrasound as a monitoring tool for natural bolides, space debris re-entries, and catastrophic launch failures.

**Highlights**

- Fine-scale atmospheric models distinguish meteoroid fragmentation from multipath.
- Simulations show discrete pulse splitting requires ranges >140 km.
- Discrete arrivals at regional ranges confirm true source fragmentation.
- Criteria enable informed monitoring of bolide and space debris re-entries.

## 1. Introduction

The interpretation of infrasound signals from energetic sources in the atmosphere is fundamentally constrained by the dynamic properties of the propagation medium. It is well established that infrasonic wavefields are strongly modulated by anisotropic fluctuations in wind velocity and temperature associated with internal gravity waves (e.g., Chunchuzov et al., 2015; Gibson et al., 2010). These ubiquitous perturbations, characterized by vertical scales significantly shorter than their horizontal extents, induce a fine-scale (FS) layered structure within the stratosphere and lower thermosphere that effectively scatters and refracts acoustic energy (Blanc-Benon et al., 2005; Chunchuzov and Kulichkov, 2020; Drob et al., 2013; LePichon et al., 2019). At regional and long-infrasonic ranges, the interaction between acoustic wavefronts and this stratified inhomogeneity can distort waveform morphology, degrade signal coherence, and introduce attenuation distinct from that predicted by horizontally homogeneous atmospheric models (Chunchuzov et al., 2025b).

While the impact of these gravity-wave perturbations on ground-based impulsive sources is increasingly well characterized (Chunchuzov et al., 2015; Chunchuzov et al., 2011), their influence on elevated, moving sources, such as meteoroids, rocket launches, and sample return capsules, remains comparatively poorly constrained. Recent work on meteoroid acoustics has emphasized the strong sensitivity of propagation models to high-resolution atmospheric specifications (Pilger et al., 2019; Pilger et al., 2020; Pilger et al., 2013; Silber and Brown, 2014). Building on this work, Chunchuzov et al. (2025b) demonstrated that vertical oscillations in the effective sound speed can systematically modify the infrasonic signatures of meteoroids fragmenting at altitudes between 35 and 100 km. Specifically, modeling shows that meteoroids fragmenting at upper atmospheric altitudes can generate distinct thermospheric arrivals generated by reflections within the 100–120 km region. We note that such "reflections" arise from strong vertical gradients and fine-scale layering in the effective sound speed and are accompanied by refraction and scattering rather than being strictly specular. Conversely, fragmentation near 50 km typically produces an acoustic shadow zone at ranges of ~150 to 300 km. Importantly, however, weak but long-lasting arrivals may still appear within this geometric shadow due to diffraction and anti-guiding (i.e., non-ducted propagation with energy leakage into the shadow zone) of low-frequency energy as the signal interacts with FS layering in the 37–45 km region. Therefore, these FS structures have been shown to shift dominant periods, elongate signal duration, and generate secondary arrivals even from a single impulsive source event. Given that the dominant signal period is a primary proxy for estimating meteoroid kinetic energy yield (Ens et al., 2012; Gi and Brown, 2017; Revelle, 1997; Silber et al., 2025d), quantifying these atmospheric contributions is a prerequisite for reliable source reconstruction.

These complex propagation effects create a fundamental ambiguity in observational meteor acoustics: a multi-arrival signature recorded at a single station may arise either from separate fragmentation events along the trajectory or from atmospheric multipath induced by FS layering (Silber and Brown, 2014). Determining which mechanism is responsible is central to reconstructing fragmentation behavior and inferring energy deposition (e.g., Borovicka and Kalenda, 2003; Brown et al., 2023; McFadden et al., 2021; Silber and Sawal, 2025). It is critical to note that this ambiguity cannot be resolved simply by correlating multi-arrival detections across spatially separated stations. Because FS atmospheric structuring is spatially heterogeneous, the propagation transfer function is unique to each source-receiver path. Consequently, the presence of multiple arrivals at one station and their absence (or difference) at another does not inherently disentangle source multiplicity from path effects. The diagnostic challenge must therefore be resolved at the level of the individual station's waveform dynamics. This distinction is particularly challenging because clear multi-arrival signatures are observationally rare in regional datasets; the vast majority of events produce a single coherent arrival, meaning that the few cases exhibiting multiple pulse trains carry disproportionate diagnostic weight.

To deconvolve these competing mechanisms, this study leverages the recently released regional meteor dataset of well-constrained events (Silber et al., 2025a; Silber et al., 2025b). This archive, comprising simultaneous observations from the Southern Ontario Meteor Network (Brown et al., 2010; Weryk and Brown, 2012) and the co-located Elginfield Infrasound Array (Edwards et al., 2008), provides a rare empirical baseline where optical trajectories, regional infrasound waveforms, and atmospheric specifications are jointly constrained at ranges less than 300 km. In this direct-arrival regime, signals typically preserve source characteristics with high fidelity, making the subset of anomalous multi-arrival events particularly diagnostic (Edwards et al., 2008; Silber and Brown, 2014).

Here we focus on event 20060305, a rare, well-observed regional meteor that exhibits clear multi-arrival structure. We employ a pseudo-differential parabolic equation (PPE) modeling framework to simulate the evolution of the acoustic field for hypothetical fragmentation sources located above, within, and below fine-scale structured layers. By systematically varying source altitude relative to these layers, we isolate the imprint of small-scale atmospheric structure on arrival coda, amplitude envelopes, and dominant periods. Comparison between modeled multipath signatures and the observed waveforms then allows us to establish physically motivated criteria for distinguishing atmospheric effects from genuine source fragmentation in regional meteor infrasound.

The paper is organized as follows: Section 2 describes the observational dataset and the selection criteria used to isolate the multi-arrival event 20060305 for detailed study. Section 3 details our approach and the pseudo-differential parabolic equation (PPE) modeling framework. Sections 4 presents simulations of acoustic field evolution for fragmentations occurring above, within, and below FS structured layers. In Section 5, we compare the modeled multipath signatures with the observed waveforms to establish physical criteria for distinguishing atmospheric effects from source fragmentation. Finally, Section 6 summarizes the findings and their implications for future monitoring of atmospheric entry events.

## 2. Dataset Description and Event Selection

The observational component of this study relies on the regional meteor dataset (Silber et al., 2025a; Silber et al., 2025b), which consists of 71 meteoroid events recorded between 2006 and 2011 by the Southern Ontario Meteor Network (SONM) all-sky optical system (Brown et al., 2010; Weryk and Brown, 2012) and the co-located Elginfield Infrasound Array (ELFO) (Edwards et al., 2008) in southwestern Ontario, Canada. This archive provides calibrated optical astrometry, reconstructed three-dimensional trajectories, synchronized infrasound waveforms, and atmospheric specification profiles for each event. It remains the only publicly available dataset globally in which optical trajectories, regional infrasound arrivals, and atmospheric conditions are jointly constrained at distances less than 300 km. This direct-arrival regime is particularly valuable because atmospheric modification is limited, enabling recorded signals to preserve source characteristics with high fidelity (Silber and Brown, 2014). A significant advantage of this dataset is that all infrasound detections originate from the same sensor array, ensuring uniform instrumentation response, identical array geometry, consistent signal processing results, and constant site noise characteristics across all events. This homogeneity eliminates inter-station variability and strengthens comparative analyses of waveform characteristics, source heights, and arrival patterns.

Of the 71 meteor events and 90 total acoustic detections, 55 events exhibit a single infrasonic arrival, and 16 events display two or more arrivals. **Figure 1** summarizes the dataset properties relevant to this study. **Figure 1a** shows the distribution of the number of arrivals per event, illustrating that multi-arrival detections constitute a minority of the dataset. **Figure 1b** presents the distribution of horizontal source–receiver ranges, all of which lie within the regional (<300 km) regime sampled by ELFO. **Figure 1c** shows the arrival-time separation for multi-arrival events as a function of range, demonstrating the diversity of multi-arrival morphology observed in the dataset. **Figure 1d** displays the distribution of mean infrasound source heights inferred from array processing, spanning roughly 40–110 km and consistent with expected ablation and fragmentation altitudes. These plots provide the observational context in which event selection must occur and illustrate that multi-arrival signals, those relevant to the present study, are intrinsically rare in the regional regime.

The rarity of multi-arrival morphology is physically meaningful and reflects the dominance of direct or quasi-direct acoustic propagation at regional distances. Only multi-arrival events contain the structure necessary to investigate whether separated arrivals arise from FS atmospheric multipath or from independent fragmentation episodes. Event selection was therefore guided directly by the structure of the dataset. All events were first screened to confirm that their detected acoustic arrivals were coherent and provided reliable backazimuth and celerity estimates. From this broader population, only those meteoroid entries that exhibited more than one clearly resolved acoustic arrival and for which the observed signals were of sufficient quality to allow meaningful comparison with model predictions, were retained for detailed analysis. This approach ensures that the events considered here possess the waveform features necessary to evaluate the potential contributions of atmospheric structure and source fragmentation to multi-arrival infrasound. The remaining single-arrival events serve as a reference, defining the expected behavior of direct-arrival meteor infrasound and constraining the natural rarity of multi-arrival signatures.

Among the 71 regional meteor events, only one case met the criteria for detailed comparison with FS propagation modeling: the 20060305 event, which exhibits two well-resolved infrasound arrivals with stable backazimuth and celerity estimates. This event provides a uniquely well-constrained observational example of multi-arrival morphology at regional distances and enables direct assessment of whether the observed arrival separation can be reproduced through FS atmospheric layering or requires multiple fragmentation episodes along the meteoroid trajectory. One additional event, 20090926, appears in this study only as a contextual example of single-arrival signals containing long-duration wavetrains; it is not used in the modeling–observation comparison. The selection of 20060305 reflects the intrinsic rarity of suitable multi-arrival observations and the need for events that provide both adequate signal quality and a resolvable arrival structure to support detailed modeling.

This dataset-driven selection strategy is appropriate for the objectives of the present work. Because multi-arrival regional meteor signals are intrinsically uncommon, only a limited number of events in any dataset, regardless of size, can provide the observational leverage required to distinguish true multi-fragmentation from atmospheric propagation effects. The curated SONM–ELFO dataset therefore serves a dual role: it establishes an empirical baseline for direct-arrival meteor infrasound under regional propagation conditions, and it provides the small but scientifically diagnostic subset of multi-arrival events necessary to evaluate the FS atmospheric propagation framework developed in this study.

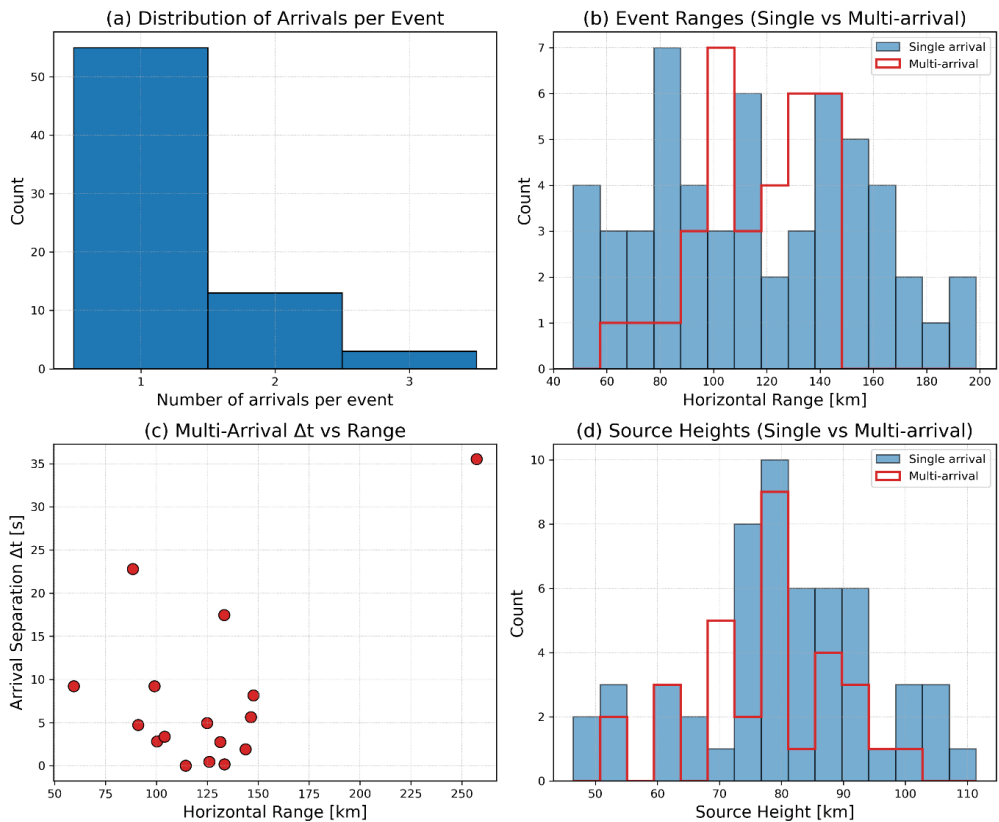

**Figure 1:** (a) Distribution of the number of infrasonic arrivals per meteoroid event recorded at ELFO. Most events exhibit a single coherent arrival, while multi-arrival cases represent a relatively small subset of the dataset. (b) Histogram of horizontal source–receiver ranges for all events, showing that detections lie entirely within the regional (<300 km) direct-arrival regime. Single-arrival events are shown as shaded bars, while multi-arrival events are overlaid as outlined bars. (c) Arrival-time separation $\Delta t$ for all multi-arrival events plotted against horizontal range. The spread in $\Delta t$ values across the sampled range interval illustrates the diversity of multi-arrival signal morphology present in the dataset and motivates the need for propagation modeling to evaluate potential contributions from source fragmentation and atmospheric structure. (d) Distribution of mean infrasound source heights derived from array processing for single-arrival (shaded) and multi-arrival (outlined) events, illustrating the sampling of 40–110 km altitudes typical of meteoroid ablation and fragmentation.

## 3. Modeling Framework for FS–Modified Meteor Infrasound

The numerical simulations undertaken here quantify how FS atmospheric structure modifies infrasound from meteoroid fragmentations at different altitudes and provide the diagnostic basis for comparison with the regional SONM–ELFO observations described in **Section 2**. The modeling integrates (i) an effective sound speed formulation that captures the combined effects of thermal stratification and wind shear; (ii) atmospheric representations containing realistic internal-gravity-wave (IGW)–driven FS structure; and (iii) forward calculations using the pseudo-differential parabolic equation (PPE) method, supplemented by ray tracing for geometric interpretation. The goal is not to reconstruct the exact atmosphere of any particular event, but to establish physically plausible propagation behaviors under realistic FS layering so that the 20060305 multi-arrival signature can be interpreted in a controlled framework.

### 3.1 Effective Sound Speed and Fine Scale Layered Structure Models

Infrasound propagation in a moving, thermally stratified atmosphere is governed by the effective sound speed, defined here as:

$$C_{\text{eff}}(z, r) = c_{\text{ad}}(z) + \mathbf{u}(z, r) \cdot \hat{\mathbf{n}},$$

where $c_{\text{ad}}(z)$ is the adiabatic sound speed $c_{\text{ad}} = \sqrt{\gamma R T(z)}$ in the absence of wind, $\mathbf{u}(z, r)$ is the three-dimensional wind field, and $\hat{\mathbf{n}}$ is the unit vector from source to receiver (e.g., Chunchuzov and Kulichkov, 2020). Vertical gradients in $C_{\text{eff}}$ therefore capture both temperature structure and the projection of horizontally sheared winds along the propagation direction, which together control refraction, ducting, and shadow-zone formation. The large-scale background atmosphere is specified using an effective sound-speed profile $C_{\text{eff},0}(z)$ derived from the Ground-to-Space (G2S) climatology, which combines empirical wind and temperature models from the troposphere to the lower thermosphere (Drob et al., 2003; Hetzer, 2024) and yields a physically realistic effective sound speed profile up to ~120 km altitude. G2S is widely used in infrasound propagation modeling and is consistent with standard atmospheric climatologies such as Horizontal Wind Model (HWM) and Mass Spectrometer and Incoherent Scatter radar (MSIS) (e.g., Antier et al., 2007; Drob et al., 2013; Drob et al., 2003; Le Pichon et al., 2008; Schwaiger et al., 2019). It provides the baseline structure into which smaller-scale perturbations are introduced.

To represent the influence of internal gravity waves, we adopt the anisotropic inhomogeneity model developed by Chunchuzov et al. (2009) and Chunchuzov et al. (2011), which prescribes vertical profiles of effective sound-speed fluctuations $\Delta C_{\text{eff}}(z, r)$ along the propagation path. In this model, IGWs with a prescribed spectrum generate small-scale perturbations in wind and temperature which are then projected into $\Delta C_{\text{eff}}$. The vertical profiles $\Delta C_{\text{eff}}(z, r)$ are computed at horizontal steps of 28 km from the source out to ~300 km and added to the background profile, yielding a two-dimensional distribution of effective sound speed along the source-receiver plane (**Figure 2**).

This IGW-based model is not arbitrary: its parameters are tuned so that the resulting vertical and horizontal wavenumber spectra of wind and temperature agree with spectra derived from independent measurements in stably stratified layers, including MST (Mesosphere-Stratosphere-Troposphere) radars (Fritts and Alexander, 2003; Tsuda, 2014), infrasound probing (Chunchuzov et al., 2013; Vorobeva et al., 2023), and high-altitude aircraft campaigns (Bacmeister et al., 1996; Chunchuzov et al., 2025a). These comparisons show that the synthetic field reproduces both the amplitude and spectral shape of mesoscale fluctuations over a wide range of scales, from ~1–5 km in the vertical to tens–hundreds of kilometers horizontally. Because the 0.8–4 Hz infrasound considered here is most sensitive to these scales, this agreement provides strong evidence that the imposed FS structure is adequate for modeling its effect on meteoroid-generated acoustic signals.

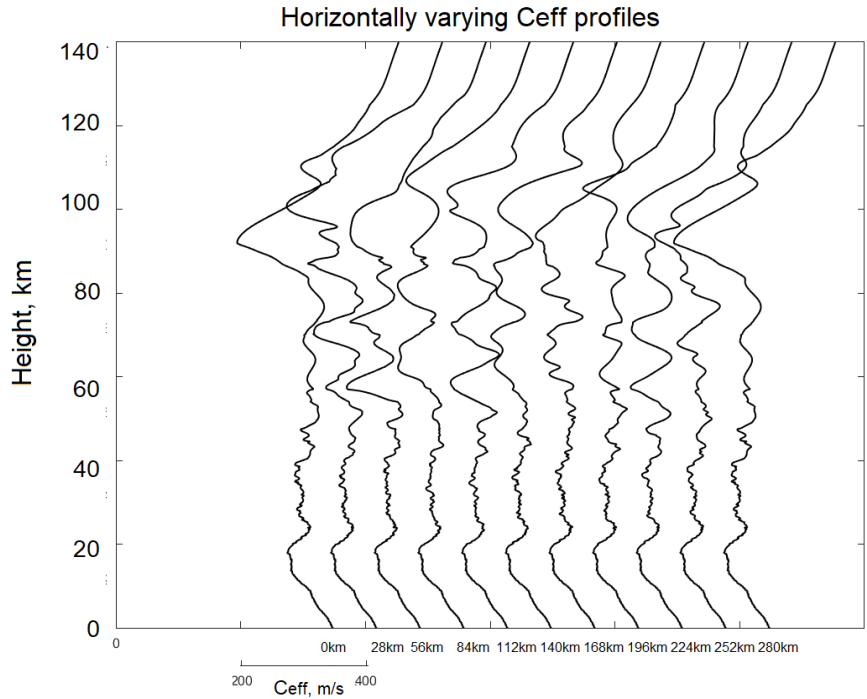

**Figure 2:** Vertical profiles of the effective sound speed *Ceff* ($z$) perturbed by anisotropic fluctuations $\Delta C_{\text{eff}}(z,r)$ which depend on the horizontal distance $r$ from the source (shown at a horizontal step of 28 km).

In addition to this fully perturbed atmosphere (Model A), we employ a diagnostic FS-layer model in which the background $C_{\text{eff},0}(z)$ is retained, but all small-scale oscillations are confined to a narrow stratospheric band between 37 and 45 km altitude (**Figure 3**). Within this layer, vertical oscillations in $C_{\text{eff}}(z)$ are imposed with amplitudes comparable to those in the IGW-based model but without horizontal variation. This idealized configuration isolates the influence of a single FS stratospheric layer on transmission and partial reflection of infrasound and is particularly relevant for regional ranges (<300 km) where an acoustic shadow zone associated with mid-stratospheric turning can form.

Therefore, our PPE calculations use two related atmospheric representations that differ only in how fine-scale structure is distributed: (i) full internal-gravity-wave (IGW) spectrum atmosphere (Model A), and (ii) isolated fine-scale stratospheric layer (Model B). Models A and B therefore serve complementary roles and bracket two end-members: a dynamically realistic IGW field spanning the entire middle and upper atmosphere, and a clean diagnostic FS stratospheric layer. Agreement between their predictions in specific altitude–range regimes is an important test of robustness; discrepancies help pinpoint which aspects of the FS field, distributed versus localized, drive particular waveform features. Since our aim is not to reconstruct the exact atmosphere of any single event but to establish physically plausible behaviors under realistic FS structure, this combination of models is adequate for the present diagnostic purpose.

By comparing simulations in these two atmospheres, we can distinguish signal modifications arising from (i) propagation through a realistic, range-varying IGW field from those caused by (ii) interactions with a single, strongly layered stratospheric region. Both representations share the same large-scale background; differences in waveform morphology therefore reflect only the presence and configuration of FS structure.

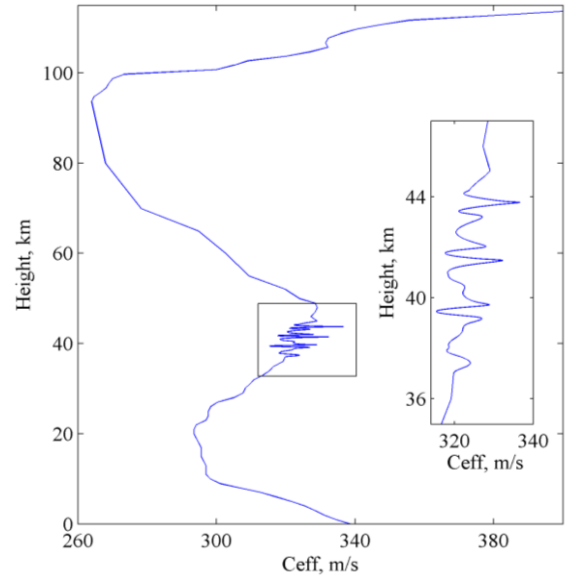

**Figure 3:** Mean vertical profile of *Ceff* (*z*) perturbed by vertical fluctuations caused by FS structure in the 37–45 km stratospheric layer.

### *3.2 Single-Fragmentation Source Representations as a Function of Source Altitude*

In the modeling, individual meteoroid fragmentations are treated as quasi-point acoustic sources. This approach is appropriate when the spatial scale of the fragmentation region is much smaller than the source–receiver distance and when the time scale of the energy release is short compared to the infrasonic periods of interest. For each experiment, the source is placed at a fixed altitude *H* within the 35–100 km range sampled by the SONM–ELFO dataset and located at the origin of the range coordinate. These source altitudes span the range sampled by regional meteor infrasound detections and allow us to characterize how FS structure modifies signals originating above, within, or below stratified layers. This stratification will be essential for interpreting regional multi-arrival observations introduced in later sections.

The source time function is chosen as a short, broadband pulse with a nearly flat spectrum between 0.8 and 4 Hz. This band is selected to match the frequency range where ELFO records meteoroid infrasound with high signal-to-noise ratio and where the PPE approximation remains valid. Near-source synthetic waveforms are stored to document the input pulse shape that would be observed in the absence of atmospheric structure. These reference signals are later used to quantify changes in waveform morphology and duration at larger ranges. Although real meteoroid fragmentation shocks can exhibit N-wave structure (Chunchuzov et al., 2025b; Silber and Brown, 2014) and mild source directivity, the present study focuses on how the atmosphere affects an initially compact pulse. Representing the fragmentation as an isotropic point source is therefore a deliberate simplification that isolates the effects of refraction, scattering, and diffraction by the FS-structured atmosphere.

### *3.3 PPE Forward Modeling and Ray Tracing*

Propagation from the source to the ground is computed using the pseudo-differential parabolic equation (PPE) method of Avilov (1995), implemented for atmospheric acoustics. The PPE solves a forward-marching approximation to the scalar acoustic wave equation in a vertical range–height plane aligned with the great-circle path between source and receiver. At each small range step, the method updates the complex acoustic pressure field using the local $C_{\text{eff}}(z, r)$ profile, allowing for continuous refraction, partial reflections at strong gradients, and scattering by FS perturbations.

The computational domain extends from the ground to at least 120 km altitude and from the source to beyond 200–250 km range, depending on the experiment. A perfectly reflecting boundary condition is applied at the ground, while an absorbing layer at the upper boundary prevents artificial reflections from the model top. Absorption in the middle and upper atmosphere is not explicitly parameterized; instead, we focus on geometric and scattering effects. In future work, frequency-dependent attenuation could be incorporated using standard absorption models, but it is not required for the diagnostic comparisons made here.

The PPE calculations are complemented by 2D ray tracing using the same $C_{\text{eff}}(z, r)$ fields. Rays are launched from the source at a set of initial elevation angles and traced through the medium using Hamiltonian ray equations. The ray solutions are used purely for geometric interpretation—identifying turning heights, shadow zones, and reflection points—and not for amplitude prediction. This dual approach leverages the strengths of both methods: ray tracing provides intuitive insight into path geometry, while PPE captures amplitude redistribution and diffraction that ray theory cannot. These tools provide a physically transparent description of regional propagation paths and arrival structure, which are the primary focus of this study. Accordingly, waveform attenuation and amplitude evolution are interpreted qualitatively in this study, while diagnostic emphasis is placed on arrival timing, separation, and propagation structure. Nonlinear, multidimensional simulations developed for reproducing detailed near-field waveform shape (e.g., Henneton et al., 2015; Nemec et al., 2017) are complementary but beyond the scope of this study.

### *3.4 Synthetic Signal Extraction and Quantitative Diagnostics*

From the PPE solutions, synthetic pressure time series are extracted at ground receivers placed at discrete ranges spanning the 70–200 km interval, which encompasses the source-receiver distances of the SONM–ELFO meteor events. For each receiver position, the complex field is inverse-transformed from frequency to time, yielding a broadband waveform. To emphasize changes in shape rather than absolute amplitude, signals are normalized by their local peak value and, where appropriate, by a $1/r$ geometric spreading factor. This normalization also facilitates comparison across different source altitudes and atmospheric models.

The synthetic signals are filtered in the same frequency band as the observations (typically 0.8–4 Hz ) and analyzed using the same diagnostics applied to the ELFO data. Time picks for the first significant peak define the direct arrival, while subsequent peaks above a prescribed fraction of the maximum amplitude are candidates for secondary arrivals. For each waveform, we compute: (i) the arrival time(s) relative to $t_0 = r/C_0$, where $C_0$ is a reference sound speed; (ii) the duration of the signal envelope, estimated from the Hilbert amplitude; (iii) the evolution of the local oscillation period, obtained from zero-crossings or short-time Fourier analysis; and (iv) the presence or absence of distinct, well-separated pulses versus a continuous "tail."

These metrics are applied uniformly across all simulations so that differences in waveform morphology can be related directly to changes in source altitude or atmospheric structure, rather than to analysis choices. In **Section 5**, the same metrics are computed for the observed ELFO signals, and the synthetic and real waveforms are compared to determine whether the observed multi-arrival signatures can be explained by fine-scale propagation alone or require multiple fragmentation events.

## 4. Altitude-Dependent Propagation Regimes in a Fine-Scale Layered Atmosphere

### 4.1 Overview of Numerical Experiments

Using the PPE framework and atmospheric models described in **Section 3**, we computed broadband acoustic fields for idealized point-like fragmentation sources at altitudes between ~35 and 100 km. Each source was embedded either in a fully perturbed atmosphere containing IGW-driven FS structure throughout 0–120 km (Model A) or in a simplified background where FS oscillations were confined to a 37–45 km stratospheric layer (Model B). Synthetic signals were extracted at ground ranges spanning ~70–200 km, encompassing the distances of the SONM–ELFO meteor events. For interpretation, it is convenient to group the experiments into three altitude regimes: (i) upper-atmosphere sources at 80–100 km, (ii) mid-stratospheric sources near 50 km, and (iii) lower-stratospheric sources below the FS layer at ~35–40 km. Within each regime we examine how FS structure influences arrival structure, waveform duration, and dominant period relative to a smooth background atmosphere.

### 4.2 Upper-Atmosphere Sources (80–100 km)

For sources placed near the top of the modeled domain (80–100 km), the PPE solutions show two distinct classes of arrivals at the ground: a direct branch, generated by rays that refract downward from the source without reflecting, and a thermospheric branch, associated with rays that propagate upward, reflect in the 100–120 km region, and then return to the surface. Meteoroid fragmentations at 80–100 km produce a clear direct arrival on the ground out to ranges of up to 400 km.

In a smooth, horizontally stratified atmosphere, the thermospheric branch emerges beyond ~140–150 km as a single coherent arrival that follows the direct branch. However, when FS fluctuations are introduced, this branch undergoes significant modification. The vertical oscillations in the effective sound speed profile create a complex reflection structure within the 100–120 km region. Ray tracing reveals that this FS layering partitions the energy into multiple paths; for example, a primary reflection may occur near 120 km, while a secondary, earlier reflection can occur near 110 km. This differential reflection causes the initially single thermospheric pulse to split into two resolvable arrivals separated by several seconds at distances around 170–180 km. (c.f., Chunchuzov et al., 2025b).

In the PPE waveforms, this splitting manifests as a compact direct pulse followed, at sufficiently large range, by one or two weaker pulses associated with different thermospheric return paths. In other words, FS structure in the 100–120 km region can turn an initially single thermospheric branch into a small family of multipath arrivals. It is important to note, however, that for the regional ranges of the SONM–ELFO dataset (<300 km), these thermospheric arrivals are often strongly damped. Our correlation analyses of events fragmenting at 80–100 km do not reveal robust thermospheric returns, a result consistent with the strong attenuation of high-frequency (>1 Hz) signal components in the 100–120 km layer (Sutherland and Bass, 2004). The modeling nevertheless indicates that multiple thermospheric arrivals could become more prominent for more energetic meteoroids whose dominant frequencies shift below ~1 Hz, where absorption is reduced and the shock wave can retain higher amplitude over longer distances.

### *4.3 Mid-Stratospheric Sources Near 50 km: Models A and B*

When the source is lowered to ~50 km, ray calculations in the fully perturbed atmosphere (Model A) show that downward-travelling rays initially reach the ground directly, but beyond a range of ~140–150 km they are bent back upward by strong vertical gradients in $C_{\text{eff}}$ near the surface. This behavior creates a classical acoustic shadow zone at the ground at ranges that depend on the details of the background profile (Chunchuzov et al.2025b, Figure 6)

Despite this geometric shadowing, the PPE solutions predict finite-amplitude signals within the shadow zone out to at least ~160–170 km. These weak arrivals arise from diffraction near near-surface caustics and from low-frequency "creeping" or anti-waveguiding modes that skim along the surface before leaking energy into the shadow region. At ranges just inside the shadow zone, the synthetic waveforms show a strong primary pulse followed by a low-amplitude, multi-cycle tail whose duration and local period increase gradually with range.

In the horizontally varying IGW atmosphere (Model A), FS structure distributed throughout the middle and upper atmosphere further modulates the shadow-zone wavefield. Perturbations near 37–45 km and again near 100–120 km act to scatter energy between nearby rays and to broaden the main arrival. Thus, for a mid-stratospheric source at 50 km, the combined effect of large-scale refraction and FS structure is to produce a single dominant arrival with an increasingly long, low-amplitude tail near and beyond the onset of the acoustic shadow zone, rather than two cleanly separated arrivals at the regional ranges of interest. The same effect of the FS structure on the signal from a source at 50 km takes place in case of atmospheric model B.

To isolate the specific influence of the 37–45 km stratospheric FS layer, we examined the 50 km source experiments using the idealized atmosphere (Model B). In this configuration, the background $C_{\text{eff}}(z)$ is smooth except for vertically oscillatory FS perturbations confined to the 37–45 km band (**Section 3**).

Ray trajectories for a 50 km source indicate that downward-propagating rays undergo total reflection in the lower atmosphere near the ground at a range of ~150 km, turning back toward the upper atmosphere. This geometry creates a classical acoustic shadow zone at the surface for ranges *r* >150–160 km. Despite this, the PPE acoustic intensity field at 1 Hz reveals nonzero energy leaking into this shadow region (**Figure 4**). This penetration is driven by diffraction at near-surface caustics, which excites "creeping waves" or anti-waveguiding modes that transport low-frequency energy along the surface into the geometric shadow (Brekhovskikh and Lysanov, 2003; Chunchuzov et al., 1990; Don and Cramond, 1986).

## Acoustic intensity calculated by PPE

### Source at a height of 50 km
### FS-structure in a height range 37-45 km

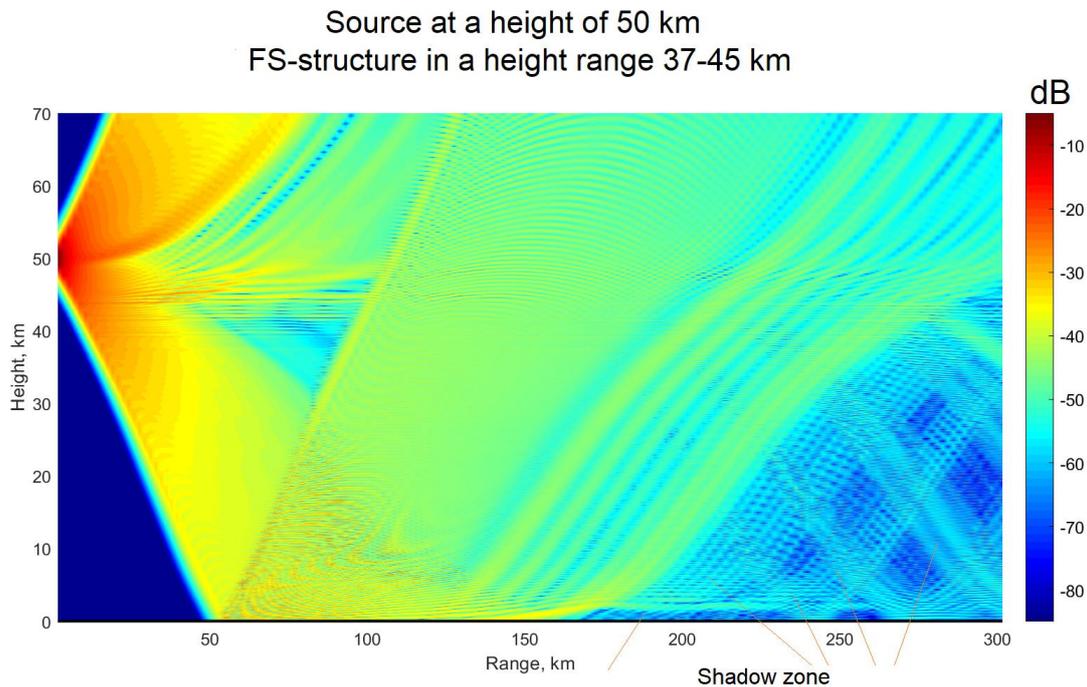

**Figure 4:** Acoustic field intensity at 1 Hz for a point source at 50 km altitude in the FS-layer atmosphere (Model B), penetrating to the geometric shadow zone.

The corresponding waveforms (**Figure 5**) demonstrate how these diffractive and scattering effects evolve with distance. At ranges $r$ <150 km, signals are dominated by a single, sharp arrival corresponding to the direct path. However, as the receiver moves into the shadow zone, the waveform morphology changes distinctively. Beyond ~150 km, a second, weaker arrival appears several seconds after the main pulse, generated by a transmission of the single pulse from a 50 km source through the stratospheric layer at 37–45 km perturbed by FS layers and antiwaveguiding propagation near ground in the shadow zone.

Importantly, the modeling allows us to quantify the "stretching" of the signal caused by these diffraction effects. As the range increases from 152 to 167 km, we observe a monotonic increase in the duration of the leading signal edge and the local oscillation period of the entire wavetrain. This range-dependent broadening is a key diagnostic of anti-waveguiding propagation. However, for the ranges most relevant to the 20060305 event (100–140 km), the 50-km source produces only a single dominant arrival with a modest trailing coda. The emergence of multiple, well-separated arrivals of comparable amplitudes is a phenomenon restricted to larger ranges (>150 km) where the shadow zone physics dominates. This will be important when we assess in **Section 5** whether the double arrivals observed for event 20060305 at ~100–107 km can be explained solely by atmospheric multipath.

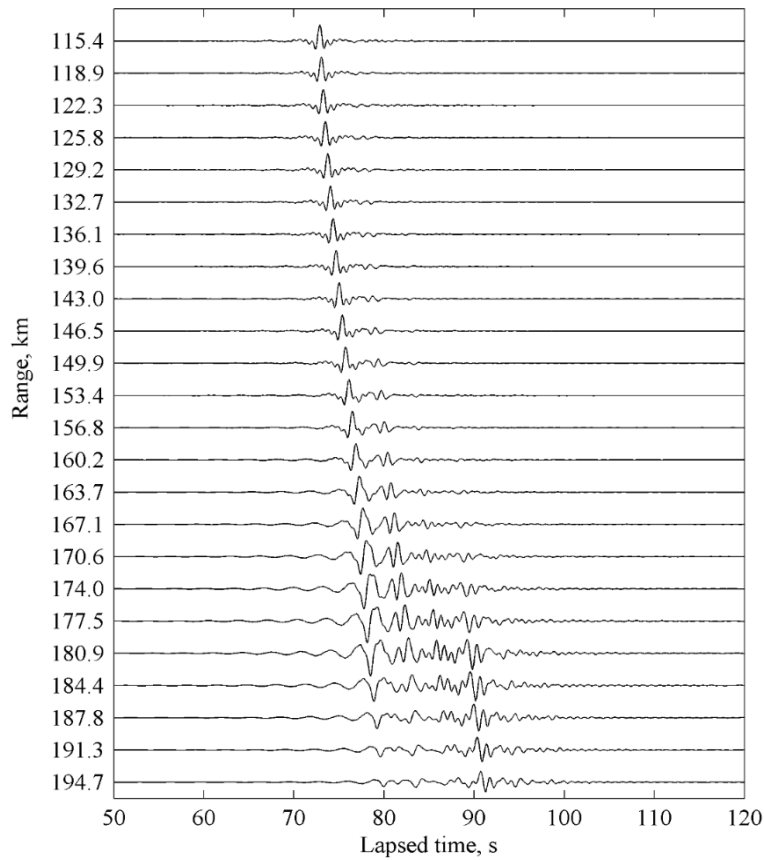

**Figure 5:** PPE-calculated signals from a source at *H* = 50 km in the FS-layer atmosphere (Model B). Lapsed time is taken relative to the moment $t_0=r/350$ m/s. At each range the signals are normalized by their maximum at this range and by the amplitude of spherical wave 1 km/*r*. Changes in the waveform, oscillation period and overall signal duration as the distance *r* increases from 136 to 167 km (PPE results).

### 4.4 Sources Below the Stratospheric FS Layer (35–40 km)

Finally, we examined sources placed below the FS layer, at ~35–40 km altitude. In this configuration, the wavefield radiated by the source must cross the overlying 37–45 km band before reaching the middle and upper atmosphere. Analytical considerations and earlier PPE studies of N-wave reflection from layered media show that such a layer can significantly stretch the reflected waveform through multiple partial reflections at different heights within the inhomogeneous zone (Chunchuzov et al., 2011; LePichon et al., 2019).

The full PPE simulations confirm this behavior. Acoustic intensity analysis for a 35 km source in the FS-layer atmosphere exhibit a near-surface caustic region where rays turn upward, again creating an extended acoustic shadow zone at ground ranges of roughly 100–220 km. Yet a horizontal slice of the PPE field at 1–1.5 Hz shows that the shadow zone is weakly "illuminated" by low-frequency energy, which leaks into the region via diffraction at the caustics and scattering within the FS layer.

The corresponding ground waveforms at ~100–140 km consist of a strong head arrival followed by a weak, oscillatory tail lasting several seconds. The head is associated with the direct or ground-reflected path, whereas the tail is generated primarily by energy that has been partially reflected within the 37–45 km band and then returned to the surface. In both the IGW-based and FS-layer atmospheres, the leading pulse remains the largest feature of the waveform, while the tail retains nearly the same back azimuth and trace velocity. No additional well-separated pulses of comparable amplitude appear in the regional range interval.

### *4.5 Summary of Altitude-Dependent Regimes*

In summary, the simulations define three distinct propagation regimes for meteoroid fragmentations in a FS-layered atmosphere:

i.  High-altitude (80–100 km) sources produce direct and thermospheric branches, with FS structure in the lower thermosphere capable of splitting the thermospheric branch into multiple late arrivals at large ranges. However, arrivals are typically attenuated below detection sensitivity.

ii. Mid-stratospheric (~50 km) sources generate a strong direct branch and, near the onset of the shadow zone ($r$ >150 km), long-duration tails and weak secondary cycles produced by diffraction and partial reflections from the 37–45 km layer.

iii. Lower-stratospheric (35–40 km) sources beneath the FS layer maintain a dominant head arrival at regional distances, but acquire extended, low-amplitude tails from partial reflections within the overlying FS band and anti-waveguiding near the ground.

Importantly, in all three regimes the appearance of clearly separated, multi-second double arrivals at ranges of ~100–110 km requires either (i) propagation through the FS stratosphere at significantly larger ranges ($\gtrsim$140–150 km), or (ii) contributions from more than one physical source. This distinction is vital for interpreting the regional observations in **Section 5**.

## 5. Comparison with Regional Meteor Observations and Diagnostic Criteria

To validate our modeling predictions against empirical 'ground truth,' we utilize specific events from the dataset (Silber and Brown, 2014; Silber et al., 2025b). While Chunchuzov et al. (2025b) established the general physical principles of FS atmospheric scattering, the present analysis extends that framework to solve a specific inverse problem: deconvolving propagation artifacts from genuine source complexity. This comparison aims to resolve the ambiguity between "diffuse" multi-arrival signatures, which our model suggests are propagation artifacts, and "discrete" multi-arrival signatures that indicate true source multiplicity.

### *5.1 "Diffuse" Multi-Arrival Event: 20090926*

We first test the propagation framework against a representative SONM–ELFO event, 20090926. This event exhibits a strong primary arrival followed by a long-duration tail defined by a continuous, oscillatory coda but it does not display a clearly resolved second arrival. As such, 20090926 provides an ideal reference case for a single fragmentation propagating through an FS layered atmosphere. The primary pulse is classified, following Silber and Brown (2014), as a Class II signal (a double-cycle N-wave), but for the purposes of this study we are primarily interested in the extended coda. We hence refer to this long-duration, low-amplitude portion of the waveform as the "diffuse" tail.

Event 20090926 corresponds to a meteoroid with a luminous path extending from 100.5 km down to 19.6 km altitude. Two signals were produced, at an altitude of ~70.8 km (range 109.1 km) and ~20.2 km (range 133.2 km). The infrasound recorded at ELFO shows a strong leading pulse followed by a modulated tail lasting ~10 s, as evident from the timeseries (**Figure 6**). After time-delay compensation and stacking across three array elements, the back azimuth and trace velocity remain nearly constant throughout both the head and tail portions of the signal. This observational feature presents a critical diagnostic challenge for real-world monitoring. In a "blind" scenario without external trajectory constraints, such a signature is intrinsically ambiguous: the constant azimuth could result from a single source dispersed by the atmosphere, or it could equally represent a sequence of multiple fragmentations occurring along a single line-of-sight azimuth. "Multi-fragmentation" events can masquerade as single prolonged signals, and conversely, single events can appear complex due to propagation.

In this specific case, however, we possess the ground truth from optical data: the signals were correlated to the altitudes of ~70 km and ~20 km. We leverage this known geometry to test whether a simplified single-source model can reproduce the observed complexity. We simulated a point source at 50 km, a representative altitude selected to capture the downward transmission through the 37–45 km FS layer. This choice also serves a practical diagnostic purpose: in many real-world monitoring scenarios, accurate trajectory constraints from optical or radar networks are unavailable, leaving infrasound as the sole diagnostic tool. Therefore, it is critical to test whether a simplified point-source model at a representative altitude can successfully reproduce the key morphological features of the signal without requiring a fully reconstructed trajectory.

The PPE waveforms at ~140–150 km range reproduce this morphology: a dominant direct arrival followed by weaker, trailing oscillations whose duration and period increase with range. The synthetic coda is generated by a combination of transmission through, and partial reflection from, the FS stratosphere, together with diffraction into the near-surface shadow region, in agreement with the qualitative behavior seen in the real data.

In principle, when meteoroids fragment below the stratopause, the layered inhomogeneities can extend and modulate the resulting infrasound wavefield, causing multiple wave arrivals and longer overall signal durations at ground-based receivers. This event demonstrates that long-duration, diffuse multi-arrival wavetrains can be fully explained by a single fragmentation episode propagating through a FS-structured atmosphere. In such cases, the presence of a tail alone is not diagnostic of multiple fragmentation; instead, it might reflect atmospheric multipath and scattering, primarily in and above the 37–45 km layer.

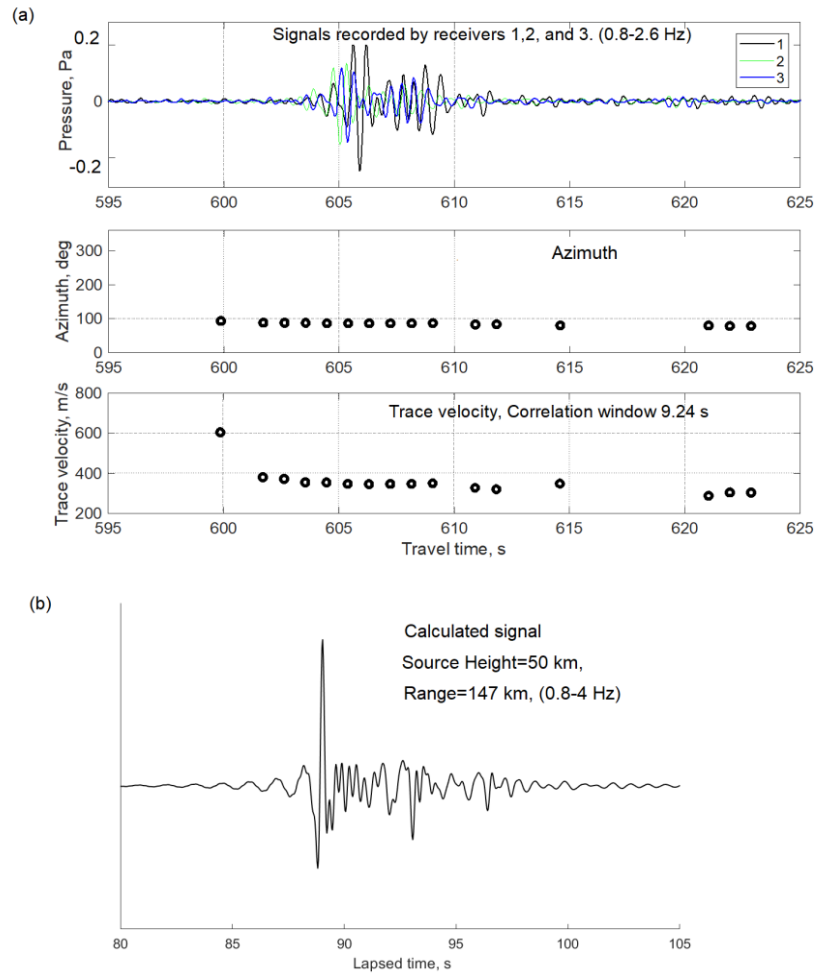

**Figure 6:** (a) 20090926 average signal obtained after time-delay compensation at receivers 1-2-3, showing azimuth and trace velocity as a function of time. (b) Signal waveform from PPE calculations in the 0.8– Hz range at *r*=147 km for a source at 50 km altitude.

### *5.2 Distinguishing Multiple Fragmentations: The Case of 20060305*

### *5.2.1 The 20060305 Event: Observations*

The 20060305 event is exceptional in the SONM–ELFO dataset because it exhibits two clearly separated arrivals at regional distances, rather than a single pulse with an attached tail. Optical analysis shows that the meteoroid experienced at least two fragmentation episodes: one near ~55 km and a closely spaced pair near ~38–39 km, all at horizontal ranges of ~100–107 km from ELFO.

Filtered infrasound recordings from three ELFO channels (2-3-4) reveal two pulses separated by ~8–10 s. Correlation-based array processing yields a back azimuth of ~347° for the first arrival and ~349° for the second, with stable celerity estimates for each (**Figure 7**). The amplitude of the second pulse is comparable to, or slightly smaller than, that of the first, and the two arrivals are separated by a period of relatively low amplitude rather than by a smooth, continuously decaying coda. This morphology is qualitatively different from the diffuse tail of 20090926 and raises the central question of this study: can such a double-arrival pattern at ~100–107 km range be produced by propagation through a FS-layered atmosphere from a single fragmentation, or does it require multiple physical sources along the trajectory?

We explore this question in the next sections.

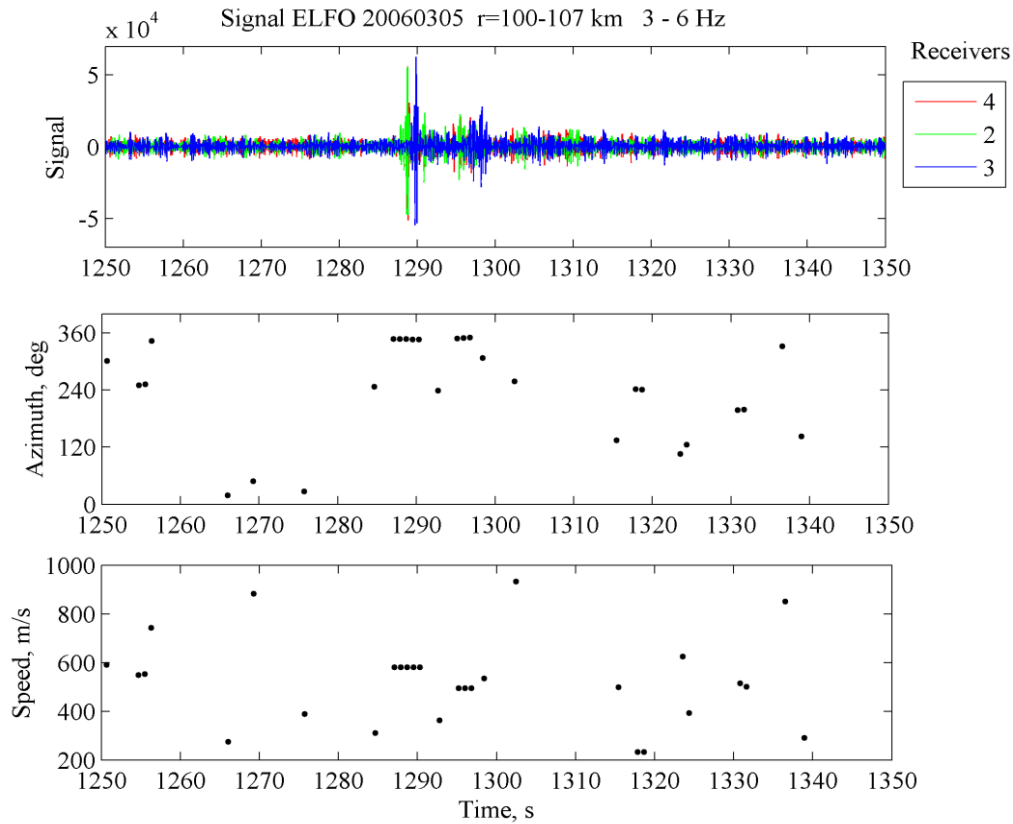

**Figure 7:** Signals (20060305) from meteoroid fragmentations at altitudes of ~55 km and 38 km, recorded by three receivers 4-2-3 at horizontal ranges *r*= 100-107 km (upper panel), their azimuth (second panel) and trace velocity (designated as speed) vs time (bottom panel).

### 5.2.2 Single-Fragmentation Scenario for 20060305

We first test the hypothesis that the two arrivals in 20060305 are generated by atmospheric multipath from a single fragmentation. To do this, we place a point source at 50 km in the FS-layer atmosphere (Model B), with $C_{eff}$ profiles representative of the event time, and compute PPE waveforms over 70–200 km. We then repeat the experiment with the source at 38 km, beneath the FS stratosphere. The corresponding ray tracing outputs for the point sources at $H$=38 km and $H$=50 km are shown in **Figure 8**.

For the 50 km source, the simulations reproduce the behavior described in **Section 4**: a strong primary arrival at ranges up to ~150 km and a gradually developing tail as range approaches and enters the acoustic shadow zone. Splitting of the primary pulse into two well-resolved arrivals of comparable amplitude occurs only at distances larger than ~140–150 km, where the combination of ground and FS reflections in the 37–45 km layer creates multiple distinct branches. At 100–110 km, the actual geometry of 20060305, the PPE waveforms show only a single dominant pulse plus a weak, closely attached coda (**Figure 9**).

The 38 km source behaves similarly: the leading arrival remains a single pulse at ranges of ~100–110 km, with an extended tail arising from partial reflections within the overlying FS layer. Even in the fully perturbed IGW atmosphere, where FS structure can scatter energy into additional paths, the simulations do not produce two strong, temporally separated pulses at these distances (**Figure 9**). Splitting again appears only at larger ranges, where the geometry favors multiple bounces between the ground and the FS layer. These results indicate that, under the atmospheric conditions considered here, a single fragmentation at either ~50 km or ~38 km cannot reproduce the observed double-arrival structure at ~100–107 km solely through atmospheric multipath. The range dependence of splitting in the model is inconsistent with the observed geometry of 20060305.

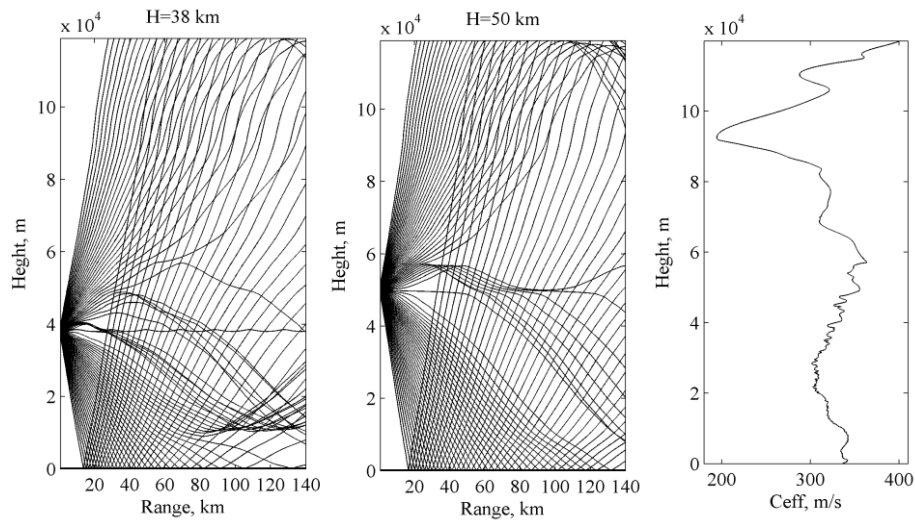

**Figure 8:** Ray trajectories from the point sources at *H*=38 km and *H*=50 km corresponding to the effective sound speed profile *Ceff* (right) with the added FS-structure in the stratospheric layer within height range 37–45 km.

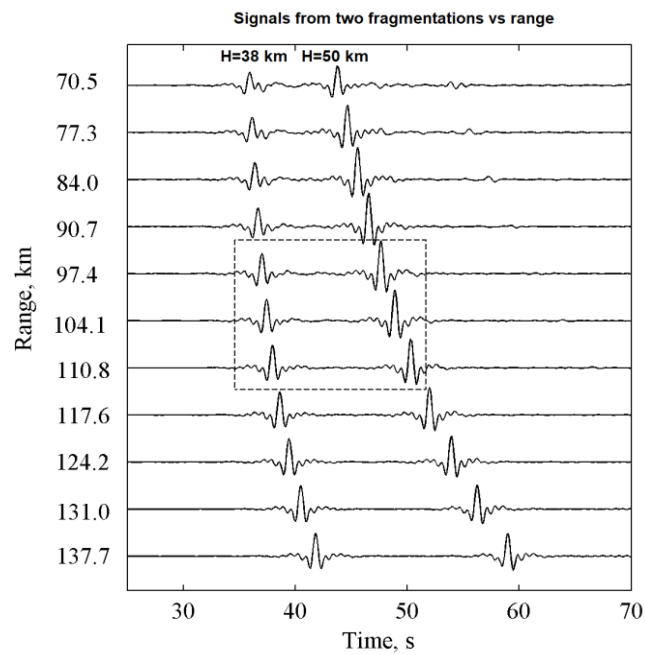

**Figure 9:** Signals calculated by PPE method from simultaneous meteoroid fragmentations at H=50 km and H=38 km vs horizontal range *r* from 70.5 to 137.7 km. The arrivals at ranges *r*=97.4-110.8 km, within which the signals in Fig.8 were recorded, are shown within a rectangle. The amplitudes of all the calculated arrivals were normalized by their maximum and by the amplitude of spherical wave 1 km/*r*. The time is taken relative to the moment $t_0$=r/340 m/s.

### *5.2.3 Two-Fragmentation Scenario for 20060305*

We next consider a model in which the 20060305 double arrival is produced by two distinct fragmentation episodes, one near 50 km and one near 38 km, occurring nearly simultaneously. Based on the reconstructed trajectory and a mean meteoroid speed of ~17 km/s, the time delay between these fragmentations is on the order of 1 s, negligible compared with the 8–10 s separation between the observed arrivals. It is therefore reasonable to treat the two fragmentations as co-temporal for propagation purposes.

In the modeling, we place two-point sources at 50 and 38 km along the observed path, inject identical broadband pulses at each, and propagate them independently through the same FS-layer atmosphere using the PPE method. The resulting synthetic signal at each ground range is the linear superposition of the two propagated wavefields.

At ranges of ~100–110 km, the simulations show one arrival from each source, with travel-time difference between the 50 km and 38 km pulses of ~8–10 s, depending on the exact range. This separation arises from differences in geometric path length and in refraction through the stratified atmosphere, not from any prescribed source time delay. Within the rectangle of ranges corresponding to the ELFO geometry (~97–111 km), the synthetic double arrivals resemble the observed pattern: two distinct pulses of similar amplitude separated by several seconds, with no additional strong arrivals in between (**Figure 10**).

At larger ranges (~140 km and beyond), the simulations also produce cases where a single fragmentation at 50 km generates a double arrival through interactions with the FS layer, but these occur only at distances significantly greater than those of 20060305, consistent with the results of **Section 4**. Spectrograms of these model signals show that the two branches at large range share similar dominant periods but are associated with different path families.

Therefore, the modeling strongly supports the interpretation that the double arrival in 20060305 at ~100–107 km is produced by two distinct fragmentation episodes at different altitudes, rather than by propagation-induced splitting of a single shock pulse.

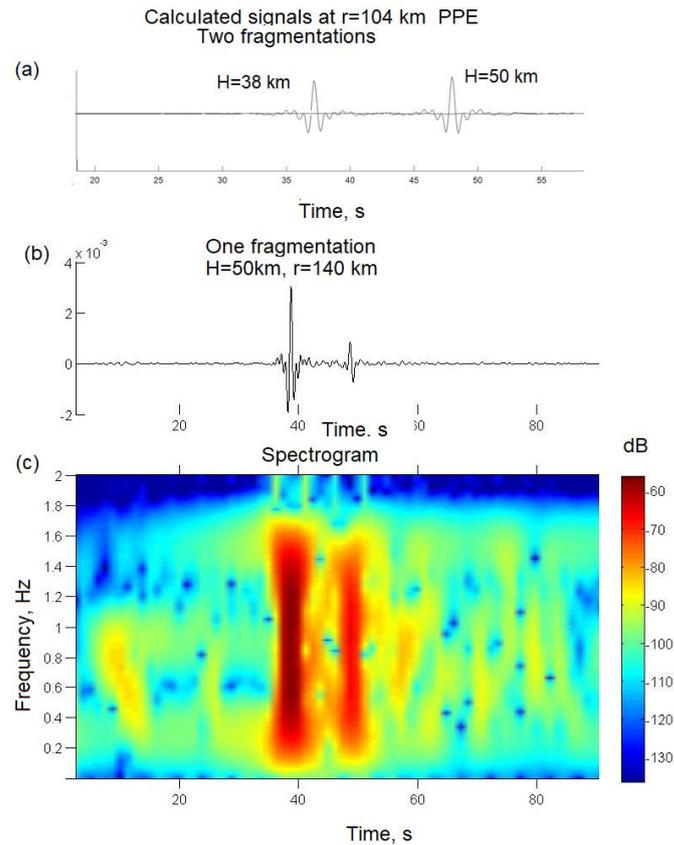

**Figure 10:** The signals calculated by PPE method at a horizontal range $r$=104 km from a source and generated by two meteoroid fragmentations at heights $H$=50 km and $H$=38 km (a); calculated arrivals of the signal at a range $r$=140 km from a source generated from one fragmentation at a height $H$=50 km (b), and spectrogram of the arrivals (c). The time is taken relative to the moment $t_0$=$r$/340 m/s.

### 5.3 Diagnostic Criteria for Distinguishing Atmospheric Multipath from Multiple Fragmentation

The combination of modeling and observations suggests several practical criteria for distinguishing multi-arrival meteor infrasound generated by atmospheric multipath from that generated by multiple fragmentation events:

i.  Range dependence of splitting: For a mid-stratospheric source at ~50 km, the PPE simulations show that significant splitting of a single pulse into two arrivals of comparable amplitude due solely to FS structure occurs only at ranges >140–150 km. At regional distances around 100–110 km, the same source produces a single dominant arrival with a trailing tail. Thus, a well-resolved double arrival at ~100 km is unlikely to be produced by a single fragmentation plus FS multipath alone.

ii.  Waveform morphology, "tail" versus "gap": Atmospheric multipath in a FS-layered atmosphere tends to produce a strong head arrival followed by a slowly decaying, often oscillatory tail (as in 20090926). The transition from head to tail is gradual. By contrast, multiple fragmentation episodes, as inferred for 20060305, produce discrete pulses separated by an interval of relatively low amplitude or noise.

iii.  Back azimuth and celerity stability: For diffuse tails, the back azimuth and trace velocity of the late-time energy closely match those of the main pulse, consistent with propagation along the same effective path. In 20060305, the two arrivals have slightly different but stable back azimuths (~347° vs. ~349°), reflecting the different altitudes and path geometries of the two sources. Differences in celerity between arrivals, if robust, also favor multiple sources.

iv. Altitude and range consistency: If the optical or radar trajectory indicates fragmentations at well-separated altitudes, the travel-time differences predicted by PPE and ray tracing for those altitudes can be compared directly with the observed arrival separation. Agreement, as in 20060305, is strong evidence for multiple fragmentation. Conversely, if all identified fragmentations occur near a single height, but double arrivals appear only at large ranges, atmospheric multipath becomes the more likely explanation.

v. Spectral evolution: Passage through FS structure generally lengthens the waveform and shifts the dominant period toward lower frequencies with increasing range, particularly in shadow zones. Multiple fragmentations can produce pulses with distinct spectral content, but in the 0.8–4 Hz band these differences may be subtle. Joint consideration of spectral evolution, range, and arrival structure helps to separate atmospheric and source contributions.

### *5.4 Implications for Meteoroid Fragmentation and Energy Estimation*

The results presented here demonstrate that FS atmospheric structure can strongly modify meteoroid-generated infrasound, producing diffuse multi-arrival signatures that, if not modeled, may be misinterpreted as multiple fragmentation events. At the same time, the 20060305 case shows that when double arrivals appear at relatively short regional ranges and are separated by several seconds, the most plausible explanation is often true source multiplicity, not purely atmospheric multipath.

Because dominant period and waveform duration are primary inputs to empirical period–yield relations used to estimate meteoroid kinetic energy (Ens et al., 2012; Gi and Brown, 2017; Revelle, 1997; Silber et al., 2025d), failing to account for FS-induced broadening can bias energy estimates, especially for signals recorded in or near acoustic shadow zones. Conversely, attributing all multi-arrival structure to fragmentation can over-interpret the source complexity. The combined PPE–observation analysis presented here provides a pathway for deconvolving these effects and for improving the robustness of fragmentation height and energy retrievals from regional infrasound.

A natural extension of this work is to embed the FS-layered propagation framework into automated analysis pipelines for regional meteor datasets and to apply the same diagnostic criteria to re-entry events of spacecraft and orbital debris, which share similar hypersonic shock-generation physics.

### *5.5 Implications for Space Debris Monitoring and Future Space Missions*

The findings presented here have profound implications for the monitoring and characterization of high-energy atmospheric events in an era of rapidly intensifying orbital activity (Pelton and Ailor, 2013). As the frequency of commercial space launches and the population of active satellites continue to grow, the inevitable consequence is a corresponding increase in re-entry events, both controlled and uncontrolled (e.g., Choi et al., 2017; Ott and Bonnal, 2025; Sawal and Silber, 2025).

A central challenge in planetary defense is discriminating between natural meteoroids and artificial objects. Space debris often produces "hybrid" shock wave signatures that lie between the "clean" ballistic N-waves of a sample return capsule (SRC) and the "chaotic" signals of a fragile or fragmenting meteoroid (Silber and Bowman, 2025; Wilson et al., 2025). By contrasting the rigorous baseline for non-ablating entry established by the OSIRIS-REx (Origins, Spectral Interpretation, Resource Identification, and Security–Regolith Explorer) (Francis et al., 2024; Lauretta et al., 2017) geophysical observational campaign (Silber et al., 2024; Silber et al., 2025c) with the discrete fragmentation signatures identified here, we provide a diagnostic framework to classify these ambiguous events. For instance, a re-entering rocket body generating a discrete multi-arrival signature with azimuthal shifts could be identified as undergoing structural breakup rather than the diffuse multi-pathing of a coherent object.

The recent acoustic characterization of the SpaceX Starship integrated flight test explosion demonstrates the value of infrasound for evaluating catastrophic vehicle failures (Bowman et al., 2025). When telemetry is lost or optical data is obscured, infrasound can independently constrain the timing, altitude, and energetic yield of the failure. The propagation regimes defined in this study, specifically the behavior of signals in the acoustic shadow zone, are directly applicable to launch and re-entry corridors (e.g., Pilger et al., 2021). Understanding how FS atmospheric structure scatters energy into these "silent" zones is critical for assessing ground impact risks and environmental noise compliance for heavy-lift launch vehicles.

Finally, the methodologies refined here apply directly to future planetary missions. The diverse, rapidly deployable sensor network used for OSIRIS-REx SRC demonstrates a viable model for instrumenting entry probes on other worlds (Fernando et al., 2025; KC et al., 2025; Silber et al., 2024; Silber et al., 2025c). On planets with thick atmospheres like Venus or Titan, or dynamic atmospheres like Mars, infrasound sensors could serve as important diagnostic tools for characterizing both vehicle performance and atmospheric structure (e.g., Bowman, 2021; Daubar et al., 2023; Dong et al., 2026; Krishnamoorthy and Bowman, 2023). The ability to deconvolve source altitude from signal period allows a single probe to vertically profile an extraterrestrial atmosphere's density and wind structure during descent, maximizing scientific return.

## 6. Conclusions

This study investigated how FS vertical layering in the middle atmosphere modifies infrasound from fragmenting meteoroids and established criteria to disentangle propagation effects from true source multiplicity. Using a PPE forward model with realistic effective sound speed fluctuations, we simulated propagation from point-like sources through internal gravity wave-driven structures in the lower thermosphere (100–120 km) and stratosphere (37–45 km).

Our results demonstrate that FS atmospheric layering predictably alters waveform morphology. Specifically, mid-stratospheric sources generate single arrivals with diffuse, oscillatory tails at regional distances, while distinct pulse splitting occurs only at ranges exceeding 140 km. By comparing these models with the 20060305 meteor event, we confirm that its discrete double arrival at ~100 km range cannot be explained by atmospheric multipathing and must instead arise from two separate fragmentation episodes. These findings provide a robust physical basis for distinguishing environmental artifacts from source physics, a critical capability for validating planetary defense models, characterizing controlled re-entries, and monitoring energetic events in an increasingly crowded near-Earth environment.

**Funding**

This work was partially supported by the Russian Science Foundation (RSF) grant № 25-17-00060 (Sections 3- 4). EAS acknowledges support by the Laboratory Directed Research and Development (LDRD) program (project number 229346) at Sandia National Laboratories, a multimission laboratory managed and operated by National Technology and Engineering Solutions of Sandia, LLC., a wholly owned subsidiary of Honeywell International, Inc., for the U.S. Department of Energy's National Nuclear Security Administration under contract DE-NA0003525.

**Acknowledgements**

This article has been authored by an employee of National Technology & Engineering Solutions of Sandia, LLC under Contract No. DE-NA0003525 with the U.S. Department of Energy (DOE). The employee owns all right, title and interest in and to the article and is solely responsible for its contents. The United States Government retains and the publisher, by accepting the article for publication, acknowledges that the United States Government retains a non-exclusive, paid-up, irrevocable, world-wide license to publish or reproduce the published form of this article or allow others to do so, for United States Government purposes. The DOE will provide public access to these results of federally sponsored research in accordance with the DOE Public Access Plan https://www.energy.gov/downloads/doe-public-access-plan. This paper describes objective technical results and analysis. Any subjective views or opinions that might be expressed in the paper do not necessarily represent the views of the U.S. Department of Energy or the United States Government, and no official endorsement should be inferred.

**Data Availability**

The full meteor event dataset is freely available on Zenodo:
https://doi.org/10.5281/zenodo.15868512.

**Conflict of Interest Disclosure**

The authors declare there are no conflicts of interest for this manuscript.